\documentclass{nature}
\usepackage[utf8]{inputenc}
\usepackage[T1]{fontenc}
\usepackage{graphicx}
\usepackage{xcolor}
\usepackage[hidelinks]{hyperref}
\usepackage{amsmath}
\usepackage{braket}

\usepackage{physics}
\usepackage{bm}
\usepackage{amsfonts}
\usepackage{amssymb}
\usepackage{upgreek}
\makeatletter
\let\saved@includegraphics\includegraphics
\AtBeginDocument{\let\includegraphics\saved@includegraphics}
\renewenvironment*{figure}{\@float{figure}}{\end@float}
\makeatother

\title{Annihilation of Dirac points and its topological obstruction in a photonic Kagome lattice}

\author{Zhaoyang Zhang$^{1\ast}$, Matthieu Finck$^{2,3}$, Changchang Li$^{1}$, Shun Liang$^{1}$,  Jerome Dubois$^3$, Yumin Tian$^{1}$, Jiahao Wen$^{1}$, Yanpeng Zhang$^{1}$, Guillaume Malpuech$^2$, Dmitry Solnyshkov$^{2,4\ast}$}

\begin{document}
\maketitle

\begin{affiliations}
    \item Key Laboratory for Physical Electronics and Devices of the Ministry of Education \& Shaanxi Key Lab of Information Photonic Technique, School of Electronic Science and Engineering, Faculty of Electronics and Information, Xi'an Jiaotong University, Xi'an 710049, China
    \item Universit\'e Clermont Auvergne, Clermont Auvergne INP, CNRS, Institut Pascal, F-63000 Clermont-Ferrand, France
    \item Laboratoire de Mathématiques Blaise Pascal, Université Clermont Auvergne, CNRS, F-63000 Clermont-Ferrand, France
    \item Institut Universitaire de France (IUF), 75231 Paris, France    
 \end{affiliations}

* marks corresponding authors, E-mails: zhyzhang@xjtu.edu.cn, dmitry.solnyshkov@uca.fr

\section*{Abstract}
\begin{abstract}
Dirac points (DPs) are topological singularities that determine the extraordinary properties of two-dimensional materials. They are generally classified by discrete topological invariants, which determine the possibility of DPs’ annihilation upon their collision. Here, we study the behaviors of DPs within a photonic Kagome lattice created in atomic vapor. With optically engineering the potential difference among three sites constituting the Kagome unit cell while preserving time-reversal symmetry and the stability of an isolated DP, the DPs move in reciprocal space. By employing conical diffraction to measure their position and the topological invariant (Euler number), we demonstrate an obstruction to DPs’ annihilation during collision and a transition to a case where the Euler number changes and annihilation occurs. Such topological transition is induced by a non-Abelian frame rotation of the eigenstates around the Brillouin zone torus. The associated conversion of the DP quaternionic charges during their motion explains the change of Euler number.
\end{abstract}

\maketitle

Topology is playing an increasingly important role in modern Physics~\cite{thouless1998topological,nakahara_geometry_2003,simon2018topology,yang2019topology}. Topological invariants with discrete values have found their practical applications in metrology for high-precision measurements~\cite{klitzing1980new,von2017metrology}. Chiral edge states are used for topological lasing~\cite{st2017lasing,bandres2018topological} and optical isolation~\cite{solnyshkov2018topological,karki2019toward,zhang2019monolithic}. Beyond that, topology provides a deeper understanding of physical phenomena, organizing possible outcomes into well-defined classes within  various types of processes. As an example, one can cite the topological defects in magnetic materials~\cite{thiaville2018topology}, liquid crystals~\cite{mermin1979topological}, quantum fluids~\cite{Pitaevskii,Zurek1985}, and even early Universe~\cite{Kibble1976}. Topological invariants are often associated with singularities~\cite{soskin2016singular,mera2021kahler}. For instance, phase singularities observed in waves and characterized by an integer winding were studied for a long time~\cite{whewell1833xi,berry2000making}, leading to advances in both classical and quantum information processing~\cite{ni2021multidimensional}.

Topology also classifies Bloch bands in solids. Beyond the well-known case of the Berry curvature of a single band and the associated band Chern number~\cite{Hasan2010}, recent works have studied band degeneracies, in particular the linear band crossings called Dirac points and the relevant invariants~\cite{montambaux_winding_2018,mergingOfDirac,lim_dirac_2020,Milicevic2019}. These works have demonstrated that in Dirac systems with more than 2 bands, accounting for multiple-band Berry curvature leads to non-Abelian behavior~\cite{wu_non-abelian_2019}. For 3-band systems with real Hamiltonians, two complementary descriptions for the topology of the Dirac points were developed, based on the Euler number and on the non-Abelian frame rotation of eigenstates evaluated via the quaternionic charges~\cite{bouhon_non-abelian_2020}. Such approaches allow predicting the possibility of annihilation of the Dirac points, or its topological obstruction~\cite{bouhon_non-abelian_2020,finck2025dirac} depending on the interplay with an auxiliary Dirac point (crossing of one of the principal bands with an adjacent band). These findings have attracted strong attention of the community~\cite{Bouhon2020,slager2024non}, and several implementations of 3-band Hamiltonians were constructed~\cite{jiang2021experimental}, as verified by experimental measurement of the corresponding band dispersions. Recently, the associated non-Abelian quantum geometric tensor has been reconstructed experimentally in a photonic system~\cite{guillot2025measuring} and a non-Abelian topological insulator has been simulated with photonic quantum walks~\cite{Lin2026}.

One of the most striking consequences of the presence of Dirac points seen as linear band crossings is the conical diffraction: the transformation of a Gaussian light beam into a hollow cone (with a ring-like cross-section). It has been observed experimentally in birefringent systems with diabolical points~\cite{Lloyd1833}, and this nontrivial result has been an important triumph of the theory~\cite{Hamilton1837}. Conical diffraction therefore allows a direct observation of Dirac points as "holes" (dark spots) in beams in real space~\cite{zhang2025transverse}. However, it is much less known that the ring formed in conical diffraction exhibits a phase singularity in the center~\cite{berry2005orbital,berry2006conical}, whose winding is determined by that of the Dirac (or diabolical) point, which is directly related to Euler number value. This feature makes it possible to probe not only the presence of the Dirac points, but even their topological invariants experimentally.

In this work, we employ a versatile photonic platform based on electromagnetically-induced transparency (EIT)~\cite{zhang2019particlelike} in atomic vapors to implement a Kagome photonic lattice, described by a real 3-band Hamiltonian exhibiting Dirac points. The reconfigurable photonic lattice is optically "written" by a Kagome-patterned coupling field~\cite{yu2022optically}. Selected lattice sites are then modulated by superimposing an additional coupling beam with a tailored periodic profile to engineer their potentials, allowing to modify the Hamiltonian. As a result, the collision of a pair of Dirac points occurs under the condition of topological obstruction preventing their annihilation. The associated Euler number is measured explicitly from the interference of the conical diffraction of a probe beam, enabling the Dirac points to be identified as phase dislocations in its interference pattern with a  Gaussian reference beam and to extract their windings. Further, we demonstrate that the Euler number describing a pair of Dirac points can be modified to remove the obstruction and allow the Dirac point annihilation. We find that this topological transition is not due to auxiliary Dirac points, but is induced by a non-Abelian frame rotation around the Brillouin zone. This frame rotation is well understood theoretically by the change of the quaternionic charges of the Dirac points, directly related to Euler number values.

\section*{Photonic Kagome lattice and its modelling}

\begin{figure}
    \centering
    \includegraphics[width=1.0\linewidth]{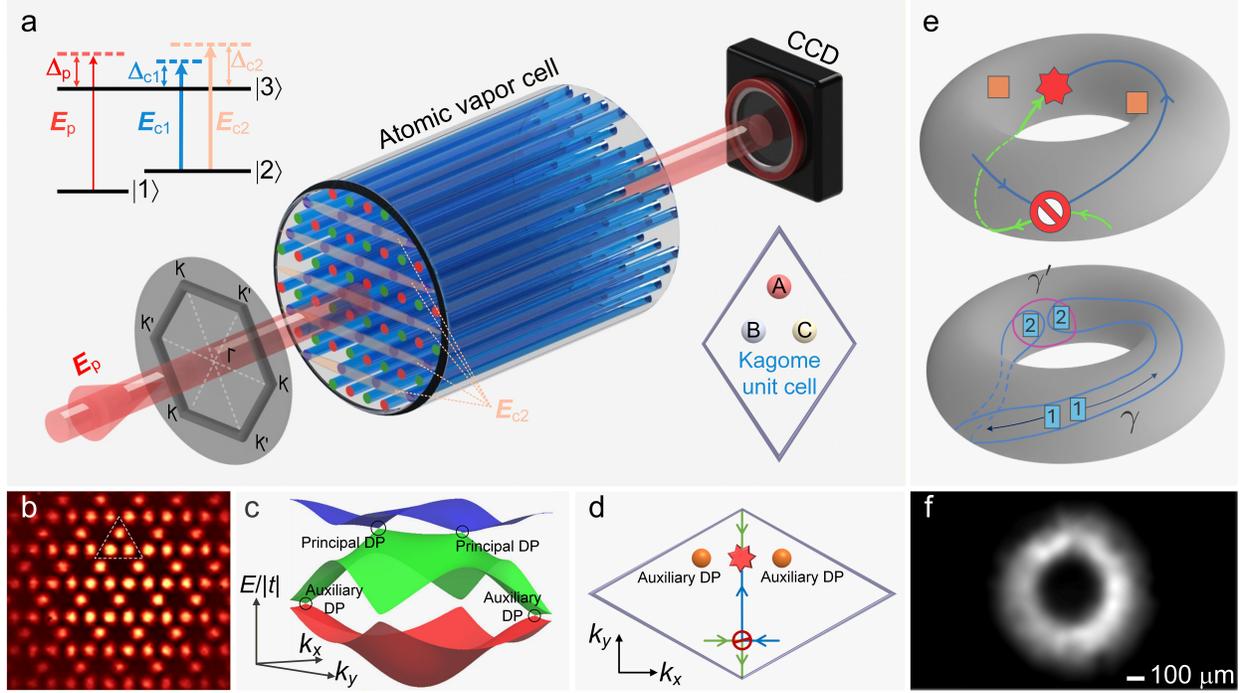}
    \caption{\small{\textbf{The Kagome photonic lattice and the Dirac points.} (a) Scheme of the experiment. Top inset: the driven atomic energy-level configuration. CCD: camera for capturing the output probe beam. Bottom inset: scheme of the lattice (unit cell with 3 sites). (b) The Kagome-patterned coupling beam for "writing" the lattice. (c) Dispersion relation $E(k_x,k_y)$ in the tight-binding approximation. (d) The 1st Brillouin zone with principal Dirac point trajectories (blue and green lines) exhibiting collision (marked as "forbidden")  and annihilation (marked with the star). The auxiliary Dirac points are marked with orange circles. (e) Top: 3D representation of the Brillouin zone as a torus, with the trajectories of the two principal Dirac points (green and blue lines) exhibiting collision with obstruction ("forbidden") followed by annihilation (star). Auxiliary Dirac points: two orange squares. Bottom: Comparison of the homotopy classes of two loops. The blue loop $\gamma$ initially encircles the two principal Dirac points at position 1 and is continuously deformed as the points move to position 2. The purple loop $\gamma'$ directly encircles the Dirac points at their final position 2, near the annihilation point. (f) Observed conical diffraction with exciting a single (auxiliary) Dirac point. }}
    \label{fig1}
\end{figure}

The experimental scheme is depicted in Fig.~\ref{fig1}(a), where the Kagome lattice is optically induced inside a Rb atomic vapor cell by a coupling field $\boldsymbol{E_{c1}}$. Its spatial distribution of intensity in the transverse plane is given in Fig.~\ref{fig1}(b), with the bright spots forming the expected Kagome lattice with in situ tunability. The other coupling field $\boldsymbol{E_{c2}}$ (with either one or two dimensional periodic distribution in the experiment as required) selectively covers the sites in the unit cell of the lattice. Both spatially periodic coupling patterns are established by a phase-type spatial light modulator with loading desired holographs~\cite{wang2024observation,liang2025observation,zhang2024nonhermitian}, see more details in Methods. The Kagome lattice can be simply described as a triangular lattice of triangles. Its unit cell with 3 sites (marked as A, B, C) is shown in the bottom right corner of Fig.~\ref{fig1}(a). Both coupling fields excite EIT for the incident Gaussian probe beam $\boldsymbol{E_{p}}$, by following a three-level $\Lambda$-type atomic configuration~\cite{feng2023loss,li2025nonreciprocal} in the top left corner of Fig.~\ref{fig1}(a). The constructed system is under the paraxial framework: the beams propagate along the $z$ direction.

In the paraxial approximation, the propagation of the probe beam is described by an equivalent of the time-dependent Schrödinger equation, where the time $t$ maps to the propagation direction $z$:
\begin{equation}
    i\hbar\frac{\partial\psi}{\partial t}=\left(-\frac{\hbar^2}{2m}\Delta+U\right)\psi
    \label{tsch}
\end{equation}
The mass $m$ of the particles is determined by dominant projection $k_z$ of the wave vector, and the potential is controlled by the atomic susceptibility $\chi$ jointly induced by the two coupling fields. Thanks to this mapping, one can use the highly efficient methods of solid-state physics to describe the beam evolution in this paraxial system, starting from the tight-binding approximation. 

The Hamiltonian of the coherently-prepared Kagome lattice in the tight-binding approximation, in the so-called base II-representation \cite{bena_remarks_2009}, reads:
\begin{equation}
H(k)=\left(\begin{array}{ccc}
E_A & t \cos \left(k \cdot \delta_1\right) & t \cos \left(k \cdot \delta_2\right) \\
t \cos \left(k \cdot \delta_1\right) & 0 & t \cos \left(k \cdot \delta_3\right) \\
t \cos \left(k \cdot \delta_2\right) & t \cos \left(k \cdot \delta_3\right) & 0
\end{array}\right)
\label{Hamiltonien}
\end{equation} 
where $E_A$ is the energy on site $A$ with respect to the other sites, $t<0$ is the tunneling coefficient between the sites, $k$ is a 2D wave vector in the Brillouin zone, and $\delta_1,\delta_2,\delta_3$ are the nearest-neighbor hopping vectors of magnitude $a$ (the lattice parameter). 
In our work, we only tune $E_A$, by setting the parameters of the extra coupling field, as we discuss below, which preserves time-reversal symmetry and the patch Euler number as a topological invariant.

The three eigenvalues $E_i(k_x,k_y)$ for $i=1,2,3$ of the Hamiltonian~\eqref{Hamiltonien} are shown in Fig.~\ref{fig1}(c) as functions of the transverse wave vector. The upper band is practically flat (completely flat in the nearest-neighbor limit for $E_A=0$). The bands exhibit several linear intersections, that we classify as principal ones (two Dirac points between the top and the medium bands) and the auxiliary ones (two Dirac points between the medium and the bottom band). Figure~\ref{fig1}(d) shows the trajectories of the principal Dirac points (blue and green arrows) in the Brillouin zone obtained by changing $E_A$ from negative to positive values. 
The collision point, where the annihilation is obstructed, is marked with the "forbidden" sign. The annihilation point is marked with a star.

The topological properties of Dirac systems can be encoded in the vector bundle formed by the eigenstates~\cite{finck2025dirac}. The annihilation can happen only when the eigenstate bundle of the two bands containing the Dirac points is topologically trivial. While the triviality of complex Hermitian bundles is typically detected by the Chern number, if the Hamiltonian is real-symmetric, the eigenstates can be set as purely real. Unlike the continuous $U(1)$ Berry phase of complex systems, the real eigenstates acquire a discrete $\pm$ phase factor after rotation around a Dirac point \cite{bouhon_non-abelian_2020}. For multi-band systems, the gauge symmetry becomes non-Abelian, meaning that the holonomy of an eigenstate along two paths in the Brillouin zone depends on their ordering. In this context, a relevant topological invariant is the Euler number \cite{ahn_failure_2019}, which can be related to quaternionic charges. 

The use of quaternionic charges and of non-Abelian frame rotation was introduced in that context in  Refs.~\cite{bouhon_non-abelian_2020}, ~\cite{wu_non-abelian_2019}, established in Ref.~\cite{bouhon2020geometric} and explained in more details in Ref.~\cite{finck2025dirac}. 
The quaternionic description is based on the fact that rotating around a Dirac point adds a $\pm$ phase on the two eigenstates of the corresponding bands. A real symmetric $3\times 3$ Hamiltonian is characterized by three orthonormal eigenvectors $u_1(k),u_2(k),u_3(k)$ forming a frame. In turn, this frame can be identified with a matrix in $O(3)$ whose columns are the eigenstates. However, each eigenstate is only defined up to a factor of $\pm1$. Consequently, the relevant space to study is the  quotient space $O(3) / \mathbb{Z}_2^3$, which can be identified with $SU(2)/Q_8$ where $Q_8=\{\pm \mathbf{1},\pm \mathbf{i},\pm \mathbf{j}, \pm \mathbf{k}\}$ is the non-Abelian group of quaternions. 
The fundamental group of this space is $Q_{8}$. We can therefore define the quaternionic charge of a loop $\gamma$ as the quaternion $c(\gamma)$ of the corresponding homotopy class. The quaternion characterizes the type of Dirac point that is encircled by the loop~\cite{wu_non-abelian_2019}. Finally, the conjugation class in $Q_{8}$ of loops encircling a Dirac point is called its quaternionic charge (see Methods).

The quaternion description can be visualized on the Brillouin zone, as illustrated in Fig.~\ref{fig1}(e).
Tuning the external parameter $E_A$ allows to modify the charge of the principal Dirac points, as shown in the top part of Fig.~\ref{fig1}(e). The trajectories of the principal Dirac points are shown with a green and a blue arrows on a torus, representing the Brillouin zone. Orange squares mark the positions of the auxiliary Dirac points which do not shift much.  
The bottom part of Fig.~\ref{fig1}(e) illustrates the evolution of the quaternionic charges of the Dirac points computed using loop homotopy classes.
In the initial configuration, at the moment of the bouncing (position 1), we trace a simple loop $\gamma$ (in blue) encircling the two points. As the Hamiltonian is modified, this loop is deformed, following the Dirac points until they arrive at the final positions (position 2). Note that this continuous deformation does not alter the homotopy class of the loop. A comparison with the reference loop $\gamma'$ (purple) reveals that the quaternionic charges satisfy the relation $c(\gamma)=-c(\gamma')$. This can be demonstrated by bringing $\gamma$ on the right of the torus handle (which results in conjugation), and composing it with the vertical cycle. 
Therefore, annihilation becomes possible thanks to the sign reversal in the quaternionic charge due to the traversal of the Brillouin zone. Our result completes the description of Ref.~\cite{bouhon_non-abelian_2020}, which considered braiding of principal Dirac points with auxiliary Dirac points. Indeed, we have demonstrated that the frame rotation phenomenon can happen not only around auxiliary Dirac points, but also around the Brillouin zone itself, because it has a non-trivial homotopy type of a torus.

Although the approaches based on the quaternionic charge and on the patch Euler number yield the same result~\cite{peng_phonons_2022}, the advantage of using quaternonic charges is that one can obtain qualitative results with only geometric reasoning over loops. Furthermore, considering the interior of the loops shown in Fig.~\ref{fig1}(e) like integration patches, the resulting topological invariant identifies with the patch Euler number, which in turn can be retrieved from the winding numbers of the Dirac points, measured experimentally in our work.

Since the conical diffraction will be our main tool for the experimental study of the Dirac points, we begin with the experimental demonstration of a conical diffraction in Fig.~\ref{fig1}(f) from a single auxiliary Dirac point, which is always well isolated. We apply Fourier filtering to the images of the conical diffraction (see Methods) to remove the discrete intensity modulation due to the lattice itself and to be able to focus only on the shape of the envelope exhibiting a typical ring in this case.

\section*{Collision of the Dirac points}
\begin{figure}
    \centering
    \includegraphics[width=1.0\linewidth]{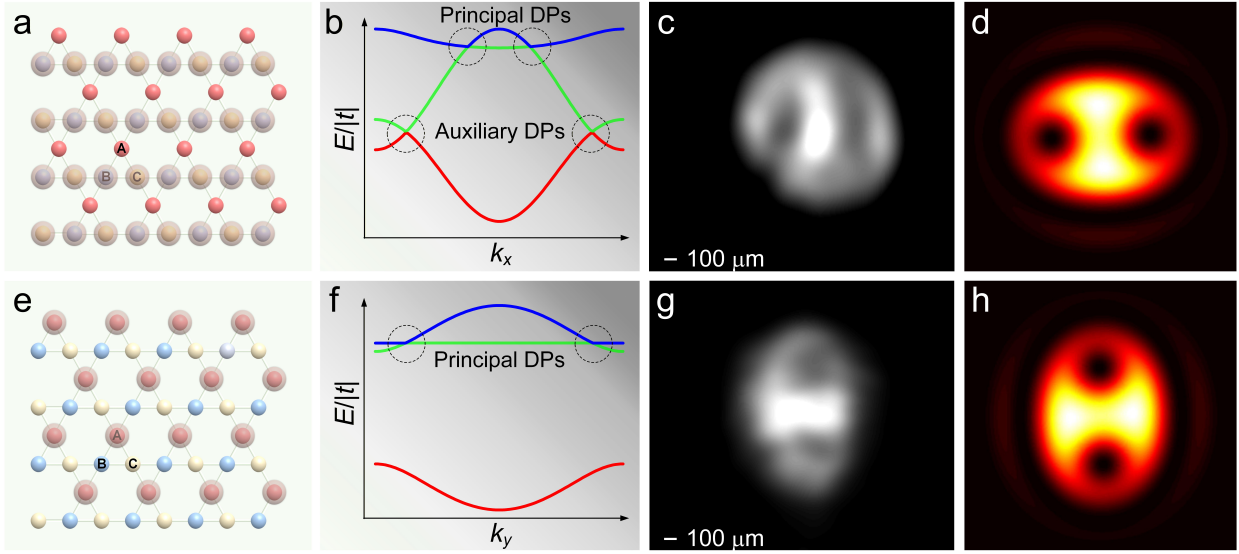}
    \caption{\textbf{Collision of Dirac points.} (a-d) $E_A<0$; (e-h) $E_A>0$. (a,e) Scheme of the experimental implementation for covering different lattice sites. (b,f) Tight-binding dispersion (along $k_x$ and $k_y$, respectively) showing the positions of the Dirac points. (c,g) Experimental images of the conical diffraction experienced by the probe showing two Dirac points along horizontal/vertical directions, respectively. (d,h) Numerical simulations of conical diffraction corresponding to $E_A<0$ and $E_A>0$, respectively.}
    \label{fig2}
\end{figure}

We now present the results of experimental measurements performed in our photonic system together with numerical simulations.
Figure 2 shows the case where the annihilation of the Dirac points is topologically obstructed. In this case, the Dirac points approach each other along $k_x$, collide (bounce), forming a 2nd-order degeneracy (parabolic band touching), and continue their trajectories along perpendicular direction $k_y$. Panels (a-d) show the configuration before the collision. Figure 2(a) shows the scheme of the additional coupling beam  covering the B and C-sites of the Kagome lattice and thus increasing their energy to meet $E_A<0$. The resulting tight-binding dispersion along the $k_x$ axis is shown in Fig.~\ref{fig2}(b). It exhibits two main Dirac points between the blue and green bands, and two auxiliary Dirac points between the green and red bands. 
We focus on the main Dirac points positioned symmetrically on either side of $k=0$, which are detected by a probe beam via conical diffraction. The incident probe beam is sufficiently broad in the $k$ space to cover both Dirac points, and its profile evolves according to \hyperref[tsch]{Eq. \eqref{tsch}}. The experimental image of the transmitted probe is shown in Fig.~\ref{fig2}(c): it exhibits two intensity minima (dark spots) along the horizontal axis. These are the images of observed conical diffraction: each intensity minimum corresponds to a Dirac point. 

To corroborate our experimental findings, we perform numerical simulations of the conical diffraction based on an effective Hamiltonian obtained with the Löwdin partitioning method~\cite{lowdin1982partitioning}. It describes the subspace of the two bands close to $k=0$ with the two principal Dirac points in the configuration where the topological obstruction prevents their annihilation. The vector space is spanned by the states in the plane perpendicular to the 3D eigenvector of the split-off band $(1,1,1)^T$. The Hamiltonian reads:
\begin{eqnarray}
    H_{eff}&=&\left(\frac{\delta}{3}-\frac{1}{2}+k_y^2-k_x^2\right)I_2\nonumber\\
    &-&2k_x k_y\sigma_x+\left(k_x^2-k_y^2-\frac{\delta}{3}\right)\sigma_z
    \label{hameff1}
\end{eqnarray}
where $\delta=E_A/t$ is the parameter of the perturbation: tuning it crossing zero allows to observe the collision of the Dirac points, which appear along $k_x$ for $\delta>0$, bounce at $\delta=0$, and reappear along $k_y$ for $\delta<0$ (we remind that $t<0$). This Hamiltonian is equivalent to the one of the optical spin Hall effect in the presence of TE-TM splitting and birefringence, well-known in photonics~\cite{terccas2014non,gianfrate2020measurement}.
The results of the simulations (see Methods for details) are shown in Fig.~\ref{fig2}(d) for $\delta>0$.

Reversing the sign of the energy of the A site with respect to the B and C sites (i. e., $E_A>0$) allows us to observe the collision of the Dirac points and the situation after the collision, when these points move along the $k_y$ axis. This inversion is experimentally achieved by reshaping the additional coupling beam to cover the A sites. The corresponding beam arrangement is shown in Fig.~\ref{fig2}(e) and the tight-binding dispersion $E(k_y)$ in Fig.~\ref{fig2}(f). The experimental conical diffraction is captured as Fig.~\ref{fig2}(g): it exhibits two dark spots along the vertical axis. They are the images of the two Dirac points, which did not disappear during the collision, because their annihilation is obstructed. The corresponding numerical simulation is shown in Fig.~\ref{fig2}(h).

 To confirm the topological nature of the obstruction, we measure the interference of the conical diffraction pattern with a Gaussian reference beam. The density minima appearing in the conical diffraction due to the Dirac points are known to have a winding of their phase determined by that of the Dirac points~\cite{berry2005orbital,berry2006conical}. It is this winding which leads to a non-trivial patch Euler number calculated on a small region of integration (patch) via the holonomy, and thus provides topological obstruction~\cite{lim_dirac_2020,bouhon_non-abelian_2020}. The difference between the patch Euler number and the winding is that the former is calculated from the eigenstates of the full Hamiltonian (3-component vectors), while the latter is calculated from the eigenstates of the effective Hamiltonian (2-component vectors). 

\begin{figure}
    \centering
    \includegraphics[width=1.0\linewidth]{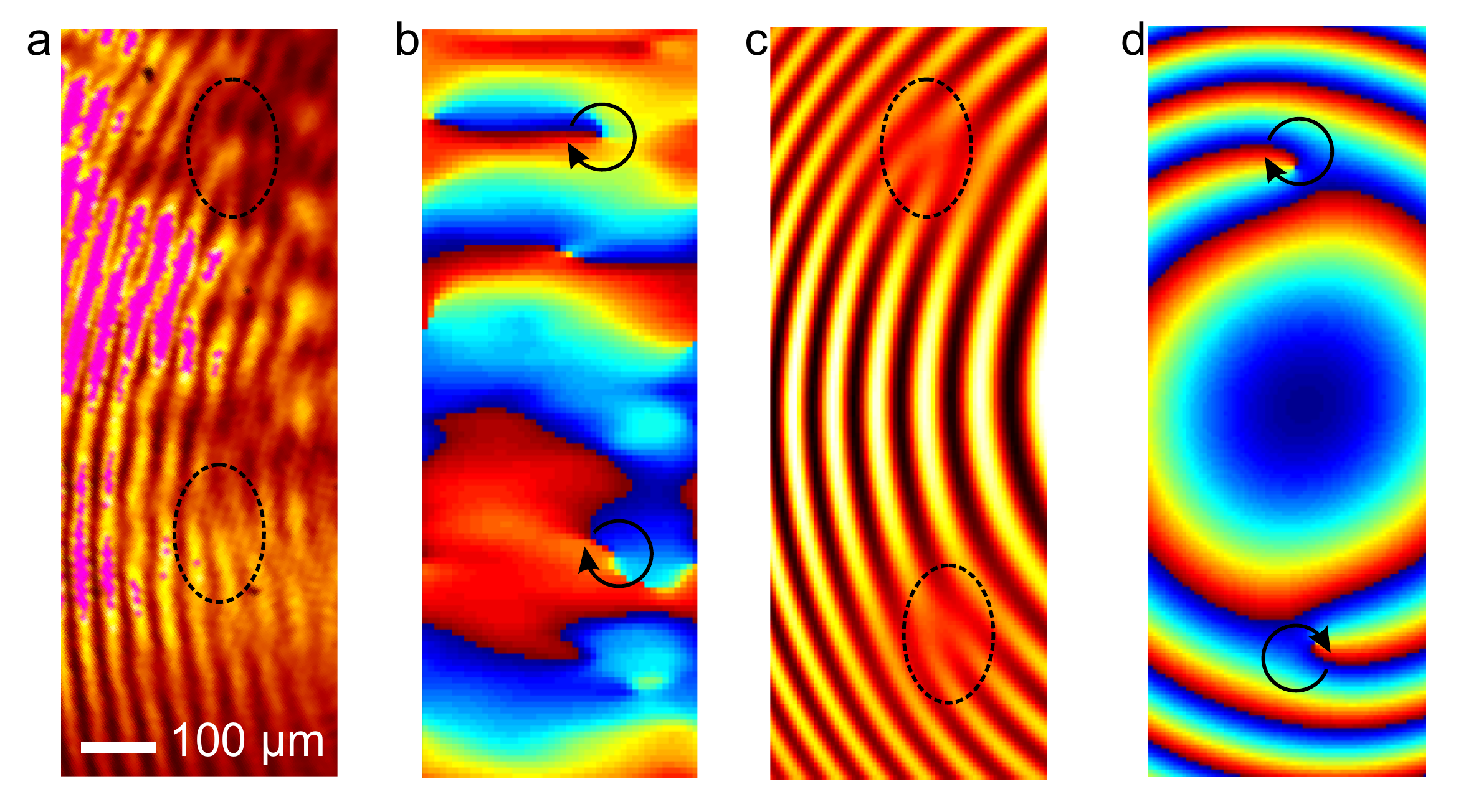}
    \caption{\textbf{Topological obstruction of the Dirac point annihilation.} (a,c) Interference pattern of the conical diffraction with a reference beam exhibiting two fork-like dislocations (experiment, theory). (b,d) Extracted phase images showing two phase singularities of the same sign (experiment, theory). }
    \label{fig3}
\end{figure}

The results of the interference measurements are shown in Fig.~\ref{fig3}(a). The pattern presents two fork-like dislocations indicated by black dashed ellipses. We extract the phase pattern from the interference via the so-called off-diagonal Fourier transform analysis~\cite{Sala2015}. The resulting phase in Fig.~\ref{fig3}(b) shows two phase singularities with the same winding (black arrows mark the phase gradient), indicating topological obstruction. The interference pattern obtained from numerical simulations and the corresponding phase are depicted in Fig.~\ref{fig3}(c) and 3(d), respectively. They exhibit the same topological obstruction as in experiment (same-sign vortices). This represents a first experimental measurement of the topological invariant (winding number and the corresponding patch Euler number) of a pair of Dirac points via conical diffraction.

\begin{figure}
    \centering
    \includegraphics[width=0.9\linewidth]{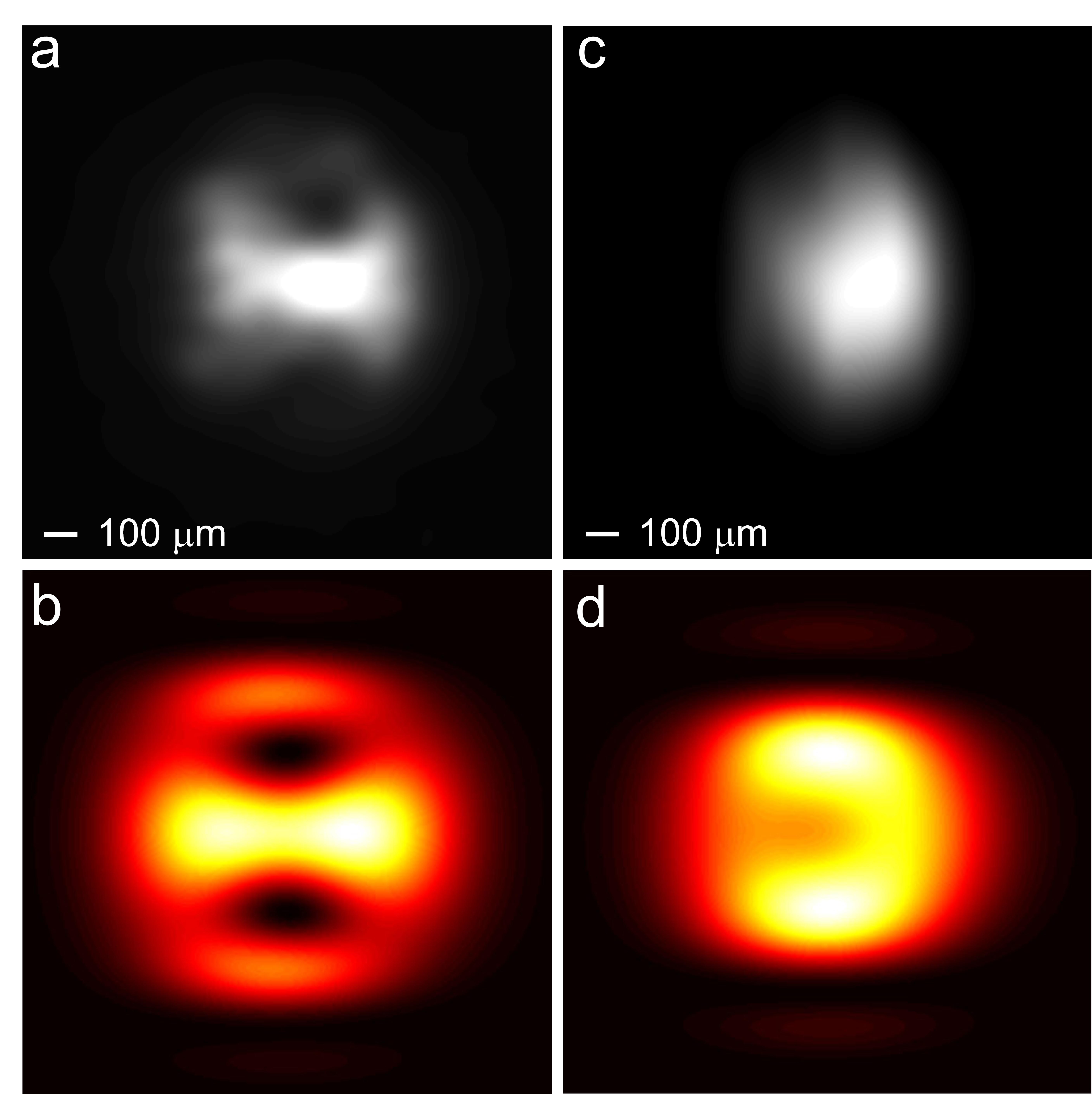}
    \caption{\textbf{Annihilation of the Dirac points.}    Experimental demonstration of two Dirac points (a) and their disappearance (c). (b) and (d) theoretically reproduced the results in (a) and (c), respectively.}
    \label{fig4}
\end{figure}

\section*{Annihilation of the Dirac points}
Finally, we focus on the annihilation of the Dirac points. The extra coupling beams are again set into a one-dimensional periodic profile to cover only A sites. The probe beam is sent to the point of the expected annihilation ($k_y\neq 0$), to detect two principal Dirac points at their new position. The observed conical diffraction appears in the vertical direction again. By controlling $E_A$ via the frequency of the $\boldsymbol{E_{c2}}$, panels (a, c) demonstrate the annihilation of the principal Dirac points [present in panel (a) and absent in panel (c)]; numerical simulations confirming this transition are shown in Fig.~\ref{fig4}(b) and 4(d). They are obtained with a different 
effective Hamiltonian valid close to the annihilation point, using the dimensionless wave vector $\bm{q}$ defined as $k_x a=q_x$, $k_y a=\pi/(2\sqrt{3})+q_y$ and defining $\delta=1+E_A/t$:
\begin{eqnarray}
H &=& \left( { - \frac{1}{2} + q_x^2 - \frac{3}{2}q_y^2 + \frac{\delta }{2}} \right){I_2}\nonumber\\
 &+& \sqrt 2 {q_x}{\sigma _x} + \left( {q_x^2 + \frac{3}{2}q_y^2 - \frac{\delta }{2}} \right){\sigma _z}    
 \label{hameff2}
\end{eqnarray}
Here, the eigenstates of the two bands are orthogonal to the 3D vector $(0,1,1)^T$, different from the case of Hamiltonian~\eqref{hameff1}.
Tuning $\delta$ though zero leads to the annihilation of the Dirac points. The textures of the eigenstates allowing to see the windings are shown in the  Supplementary Materials.

The versatile tunable photonic platform represented by atomic vapors under EIT has allowed us to study  experimentally  the behavior of Dirac points in the Kagome lattice. Using conical diffraction resulting in ring-like intensity distribution at each of the Dirac points, we have  demonstrated the "bouncing" of the Dirac points due to the topological obstruction of their annihilation. Further, we show that this obstruction is suppressed by non-Abelian frame rotation of the eigenstates around the Brillouin zone torus. This mechanism allows the annihilation to occur, and we successfully observed it in our experiment. Finally, our work opens the door to experimental measurements of topological invariants by exploiting the phase of the electromagnetic wave acquired in the  well-known conical diffraction associated with Dirac points.

\begin{methods}
\textit{Experiment.}
The Gaussian probe field $\boldsymbol{E_{p}}$ (frequency $\omega_{p}$) is slightly defocusing during its propagation, ensuring that it contains multiple wavevectors to match two target principal Dirac points with a small interval in the $K$ space. Both the required periodical coupling fields with desired patterns are produced by a spatial light modulator (with resolution being 1920×1152). Particularly, the reshaping of extra coupling field $\boldsymbol{E_{c2}}$ is easily realized by loading different holographs. The probe beam with a power of ~2 mW connects the atomic transition from ground state $\ket{1}$ ($\mathrm{5S}_{1/2},\ F=2$) to the upper state $\ket{3}$ ($\mathrm{5P}_{1/2}$) with detuning $\Delta_{p}$. Both coupling fields $\boldsymbol{E_{c1}}$ and $\boldsymbol{E_{c2}}$ ($\omega_{c1}$, $\omega_{c2}$) couple the same transition between $\ket{2}$ ($\mathrm{5S}_{1/2},\ F=3$, the other hyperfine level of the ground state) and $\ket{3}$ with detunings $\Delta_{c1}$ and $\Delta_{c2}$, see Fig.~\ref{fig1}(a). Actually, both $\boldsymbol{E_{c1}}$ and $\boldsymbol{E_{c2}}$ (with power being 30 mW and 40 mW, respectively) can produce EIT effect, but the two EIT windows do not completely overlap with each other since $\Delta_{c2} \neq \Delta_{c3}$. Here the detuning $\Delta_{p}$ ($\Delta_{c1}$, $\Delta_{c2}$) is defined as $\omega_{p}-\omega_{13}$ ($\omega_{c1}-\omega_{23}$, $\omega_{c2}-\omega_{23}$), with $\omega_{13}$ ($\omega_{23}$) denoting the frequency gap between $\ket{1}$ ($\ket{2}$) and $\ket{3}$.

The spatially modulated susceptibility is governed by the intensity and frequency detuning of involved fields sent into the adopted three-level atomic vapor cell, and its detailed expression is given in Ref. \cite{yu2022optically}. In the experiment, the probe and Kagome coupling detuning are fixed at $\Delta_{p}=130$ MHz and $\Delta_{c1}=80$ MHz. The detuning of $\boldsymbol{E_{c2}}$ is $\Delta_{c2}=105$ MHz in Fig.~\ref{fig2}, while it is set as $\Delta_{c2}=90$ MHz (conical diffraction) and $\Delta_{c2}=116$ MHz (annihilation of conical diffraction) in Fig.~\ref{fig4}. All the involved three fields are from tunable semiconductor laser sources that can emit continuous-wave light. The experiment is conducted within a 5-cm-long rubidium vapor cell, whose temperature is stabilized at 100°C using a custom-built controller.

\textit{Conical diffraction images.} In the main text, we show the conical diffraction images after Fourier-filtering, which allows to remove the periodic modulation due to the Kagome lattice itself. This allows to enhance the visibility of the minima corresponding to the Dirac points. The examples of original images are shown in Supplemental Materials.

\textit{Numerical simulations.}
We solve the time-dependent 2D spinor Schrödinger equation with the effective Hamiltonians~\eqref{hameff1},\eqref{hameff2}  using the 3rd order Adams-Bashforth method with Fourier-transform calculations  accelerated by the Graphics Processor Unit. 

\textit{Patch Euler number.} For a real symmetric Hamiltonian with three bands ordered by increasing energy $E_1 \leq E_2 \leq E_3$, the Euler connection between bands 2 and 3 is defined as $A_k = \langle u_2(k) \mid d_k u_3 \rangle$, and the Euler curvature is its exterior derivative $F = dA$. When the two principal bands are not isolated from the third one everywhere in the Brillouin zone, the Euler number is ill-defined on the whole torus, but one can still define a patch Euler number on a region $D$ that does not contain adjacent Dirac points~\cite{ahn_failure_2019,bouhon_non-abelian_2020,peng_phonons_2022}:
\begin{equation}
    e_2(D) = \frac{1}{2\pi}\left[\int_{D} F_k - \oint_{\partial D} A_k\right]
\end{equation}
In practice, the patch Euler number can be efficiently computed by complexifying the real eigenspace bundle and computing the corresponding Chern number~\cite{bouhon_non-abelian_2020}.

\textit{Quaternion conjugation classes:}
We can identify 5 quaternion conjugation classes:
\begin{itemize}
\item 1 corresponds to a loop with no Dirac points inside, or two Dirac points which can annihilate inside the loop.
\item $\pm \mathbf{i}$ corresponds to a principal Dirac point.
\item $\pm \mathbf{k}$ corresponds to an adjacent Dirac point.
\item $\pm \mathbf{j}$ corresponds to a pair of each type of Dirac points.
\item $-1$ corresponds to a pair of Dirac points of the same type and which cannot annihilate inside the loop.
\end{itemize}

\end{methods}

\section*{References}

\bibliographystyle{naturemag}
\bibliography{biblio}

\begin{addendum}
\item This work was supported by National Natural Science Foundation of China (No. 62475209). Additional support was provided by the ANR Labex GaNext (ANR-11-LABX-0014), the ANR program "Investissements d'Avenir" through the IDEX-ISITE initiative 16-IDEX-0001 (CAP 20-25), the ANR projects MoirePlusPlus (ANR-23-CE09-0033), HAWQ (ANR-25-CE47-7323), the ANR SFRI project ”Graduate Track of Mathematics and Physics” (GTMP) of University Clermont Auvergne, Qinchuangyuan “Scientist+Engineer” Team Construction of Shaanxi Province (2024QCY-KXJ-178), and Key Research and Development Program of Shaanxi Province (2025PT-ZCK-49).
\item[Author contributions] Zhaoyang Zhang – project administration, investigation, formal analysis, funding acquisition, methodology, visualization, investigation, writing; Changchang Li – investigation, visualization;
Shun Liang– investigation, methodology; Yumin Tian-investigation; 
Yanpeng Zhang – funding acquisition and formal analysis;
Jiahao Wen – investigation; Matthieu Finck - investigation, visualization, writing; Jerome Dubois - investigation, supervision; Guillaume Malpuech - project administration, funding acquisition, investigation, writing; Dmitry Solnyshkov - project administration, data analysis, investigation, visualization, writing. All authors have contributed to the manuscript and have read the final version of the manuscript.
\item[Competing interests] The authors declare no competing interests. 
\item[Correspondence] Correspondence
should be addressed to Zhaoyang Zhang (zhyzhang@xjtu.edu.cn) and Dmitry Solnyshkov (dmitry.solnyshkov@uca.fr).

\end{addendum}

\setcounter{equation}{0}
\setcounter{figure}{0}
\setcounter{table}{0}
\setcounter{page}{1}
\setcounter{section}{0}

\makeatletter
\renewcommand{\theequation}{S\arabic{equation}}
\renewcommand{\thefigure}{S\arabic{figure}}
\renewcommand{\theHfigure}{S\arabic{figure}}
\renewcommand{\thetable}{S\arabic{table}}
\renewcommand{\thesection}{S\Roman{section}}
\renewcommand{\thepage}{S\arabic{page}}

\section*{Supplementary Materials}

Here, we provide additional experimental and theoretical results.

\subsection{Effective Hamiltonian computation}
We provide here the steps of the Löwdin partitioning method~\cite{lowdin1982partitioning} explicitly.
As described in the main text, we choose a point of the Brillouin zone (for example, the collision point $k=0$) as a reference for the unperturbed Hamiltonian $H^{(0)}$. Around this point, the wavevector-dependent terms are considered to be small, they are the small perturbation $H^{(1)}=H(k)-H^{(0)}$. The subspace $\alpha$ corresponds to the two bands with the principal Dirac points, while the subspace $\beta$ corresponds to the third band (split-off). The two eigenvalues of the $\alpha$ subspace at the reference point are $E_1^{(0)}$ and $E_2^{(0)}$, while the eigenvalue of the subspace $\beta$ is $E_3^{(0)}$.
The average energy of the states of the subspace $\alpha$ is defined as $E_\alpha=(E_1^{(0)}+E_2^{(0)})/2$. Then, the terms of the effective Hamiltonian are given by
\begin{eqnarray}
\label{Lowdin}
    H^{eff}_{ij}&=&\delta_{ij}E_i^{(0)}+\bra{\psi_i^{(0)}}H^{(1)}\ket{\psi_j^{(0)}}\\
    &-&\frac{\bra{\psi_i^{(0)}}H^{(1)}\ket{\psi_3^{(0)}}\bra{\psi_3^{(0)}}H^{(1)}\ket{\psi_j^{(0)}}}{E_3^{(0)}-E_\alpha}\nonumber
\end{eqnarray}
where $i,j=1,2$ (the $\alpha$ subspace) and the state $3$ corresponds to the $\beta$ subspace.
Depending on the particular case, series expansion in terms of $k$ or $q$ can be applied to $H^{eff}$ or to $H^{(1)}$.

\subsection{Effective Hamiltonian textures}
In order to facilitate the understanding of the effective Hamiltonians presented in the main text, we show here the pseudospin textures (in the XZ plane) corresponding to both of them.

\begin{figure}
    \centering
    \includegraphics[width=0.9\linewidth]{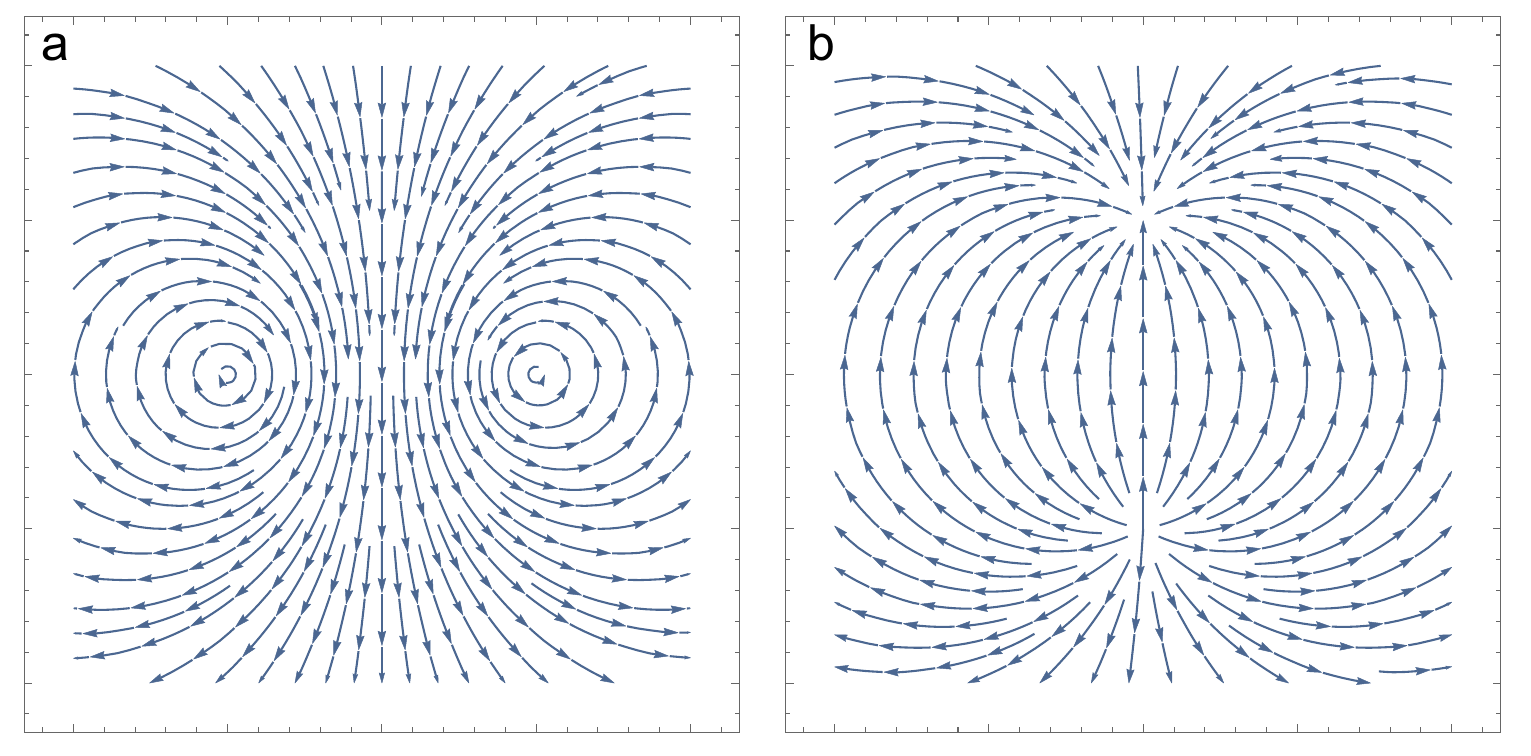}
    \caption{\textbf{Texture of the eigenstates of the Hamiltonian (3) from the main text}: \textbf{a} before and \textbf{b} after collision of the Dirac points with the same winding.}
    \label{figScollHam}
\end{figure}

Figure~\ref{figScollHam} shows the texture of the eigenstates of the effective Hamiltonian (3) of the main text. The winding of both Dirac points is the same, and therefore their collision cannot lead to their annihilation. Fig.~\ref{figScollHam}(a) corresponds to Fig.~2(c,d) from the main text ($E_A<0$, Dirac points located along the horizontal axis), whereas Fig.~\ref{figScollHam}(b) corresponds to Fig.~2(g,h) from the main text ($E_A>0$, Dirac points located along the vertical axis).

\begin{figure}
    \centering
    \includegraphics[width=0.9\linewidth]{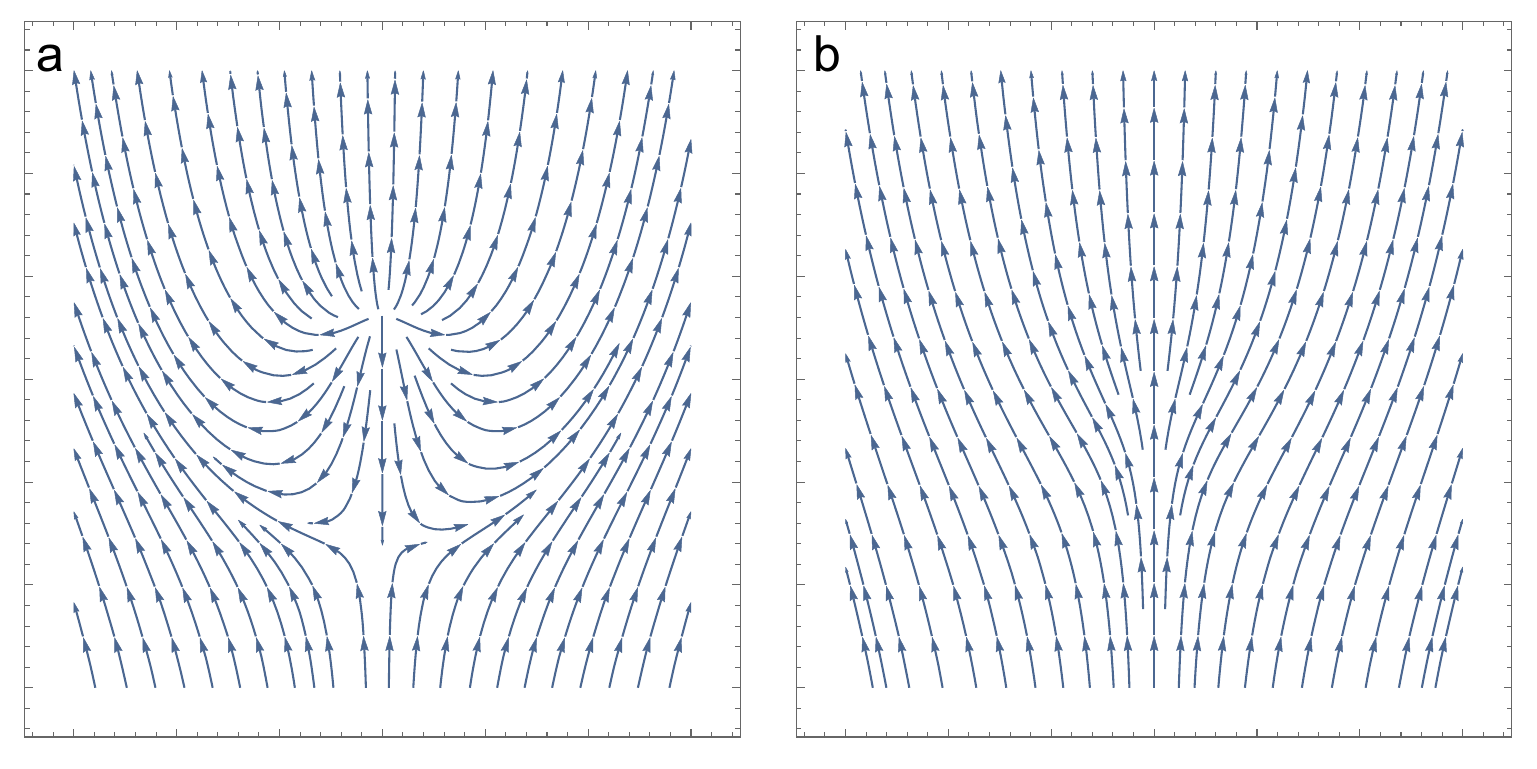}
    \caption{\textbf{Texture of the eigenstates of the Hamiltonian (4) from the main text}: \textbf{a} before and \textbf{b} after annihilation of the Dirac points with opposite windings.}
    \label{figSannihHam}
\end{figure}

Figure~\ref{figSannihHam} shows the texture of the eigenstates of the effective Hamiltonian (4) of the main text. The winding of the Dirac points is opposite [panel (a) corresponding to Fig.~4(c,d) of the main text], which makes their annihilation possible: no singularities are visible in panel (b), corresponding to Fig.~4(e,f) of the main text.

\subsection{Winding of the Dirac points}
\begin{figure}
    \centering
    \includegraphics[width=0.8\linewidth]{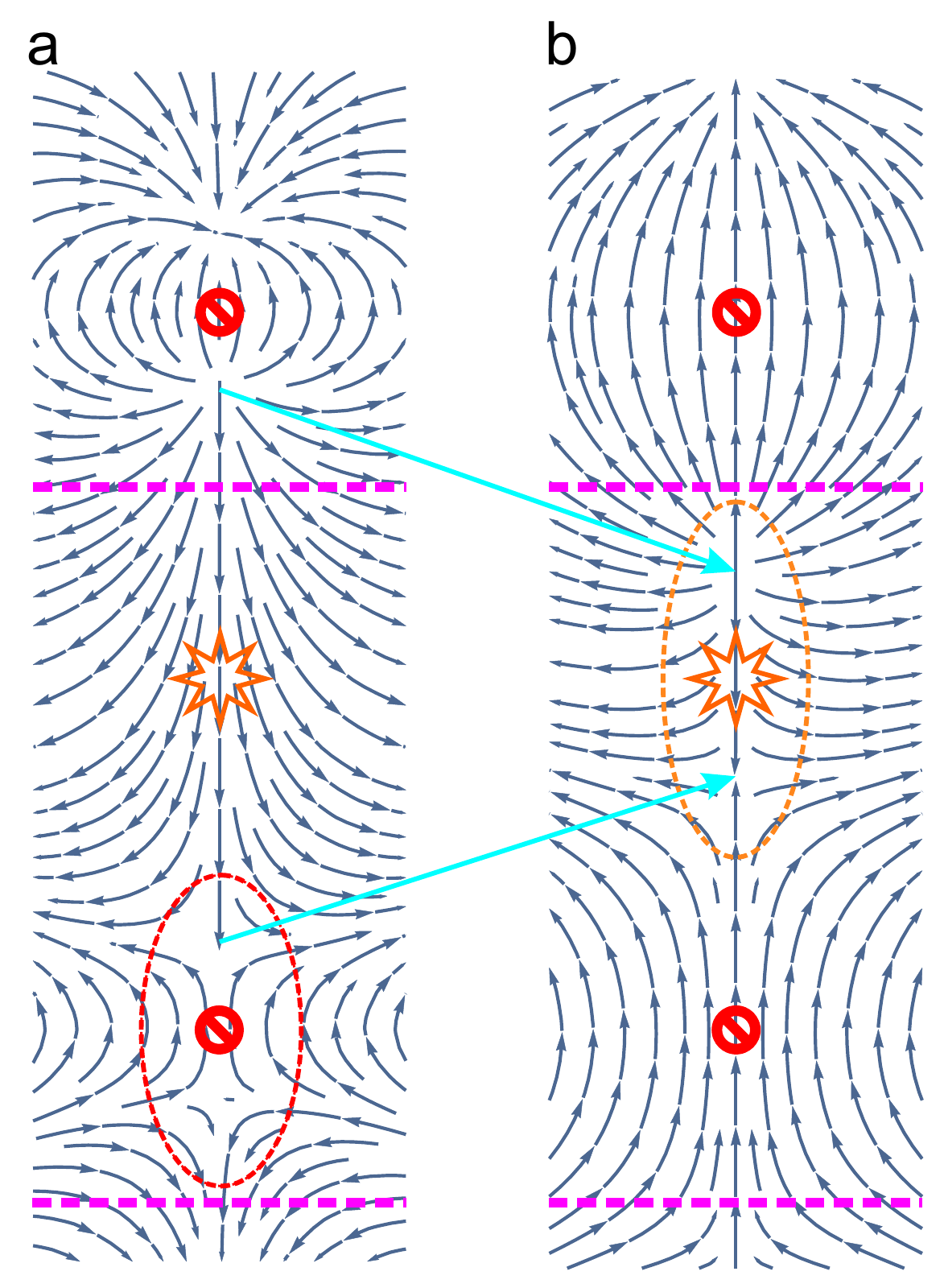}
    \caption{\small \textbf{The texture of the eigenstates of the general effective Hamiltonian~\eqref{hefflarge}.} \textbf{a} Close to the collision of the Dirac points;  \textbf{b} Close to the annihilation of the Dirac points. Dashed magenta lines mark the BZ boundaries. The "forbidden" signs mark the collision points and the "star" marks the annihilation point. The cyan arrows indicate the motion of the Dirac points of interest between the two panels. The ellipses indicate the regions with 2 Dirac points for which the topological invariants are evaluated.}
    \label{figSehala}
\end{figure}

The Kagome lattice can be described by a real $3\times 3$ Hamiltonian (2) of the main text. This Hamiltonian allows describing the transition between collision and annihilation of the Dirac points within a single effective Hamiltonian obtained with a fixed decomposition basis, valid in a narrow band close to $k_x=0$, without limitations on $k_y$. 
\begin{eqnarray}
    H_{eff}&=&\sqrt{2}\cos\left(\sqrt{3}k_y\right)\sin\left(k_x\right)\sigma_x\label{hefflarge}\\
    &+&\frac{1}{2}\left(\cos^2\left(k_x\right)\sin^2\left(\sqrt{3}k_y\right)-\cos\left(2k_x\right)-\delta\right)\sigma_z\nonumber
\end{eqnarray}
In this picture, the signs of the Dirac points never change; however, the Dirac points in adjacent Brillouin zones (BZs) have opposite signs~\cite{lim_dirac_2020}. A Dirac point that participated in a collision within a given BZ (e.g. the 1st BZ) then meets a Dirac point arriving from an adjacent BZ (e.g. the 2nd BZ), which has an opposite sign, and annihilates with it. This process is described in the main text in terms of the loop homotopy classes, and the opposite signs of Dirac points from adjacent BZs are due to the non-trivial homotopy of the BZ torus. Here, we illustrate it in Fig.~\ref{figSehala}, demonstrating the pseudospin texture in the XZ plane of the Hamiltonian~\eqref{hefflarge} for $\delta=-0.1$ [close to the collision, panel (a)] and $\delta=-0.8$ [close to the annihilation, panel (b)]. 
The Dirac points appear as singularities of the pseudospin texture: the positive winding shows up as divergent/convergent "monopoles", whereas the negative winding appears as the "Dresselhaus"-like field texture.
The edges of the 1st BZ are marked with dashed magenta lines. We only show a narrow region along $k_x$, which corresponds approximately to the validity of the Hamiltonian~\eqref{hefflarge}. The collision points are marked with red "forbidden" signs, whereas the annihilation point is marked with an orange star. The cyan arrows indicate the motion of the Dirac points of interest between the two images. The upper of these two Dirac points crosses the edge of the BZ. Finally, the dashed dark red and orange ellipses mark the loops around a pair of Dirac points, corresponding to the topological invariants described in the main text.

On the other hand, the Kagome lattice can also be described with a complex Hamiltonian: 
\begin{equation}
{H_c} = \left( {\begin{array}{*{20}{c}}
\delta &{\frac{{1 + {e^{i\left( {{k_x} + \sqrt 3 {k_y}} \right)}}}}{2}}&{\frac{{1 + {e^{2i{k_x}}}}}{2}}\\
{\frac{{1 + {e^{ - i\left( {{k_x} + \sqrt 3 {k_y}} \right)}}}}{2}}&0&{\frac{{1 + {e^{i\left( {{k_x} - \sqrt 3 {k_y}} \right)}}}}{2}}\\
{\frac{{1 + {e^{ - 2i{k_x}}}}}{2}}&{\frac{{1 + {e^{ - i\left( {{k_x} - \sqrt 3 {k_y}} \right)}}}}{2}}&0
\end{array}} \right)
\end{equation}
In this case it is impossible to choose a fixed basis, valid both at the moment of collision and at the moment of annihilation. Nevertheless, it is possible to analyze the rotation of the winding vectors of the Dirac points (with their positions given by $\delta=(\cos(2\sqrt{3}k_y)-1)/2$) close to their collision and annihilation points, and this analysis is in agreement with what could be expected from the previous works~\cite{lim_dirac_2020}: when the two Dirac points move away from the collision points, their winding vectors (which were initially aligned) rotate in opposite directions. For example, the Dirac points in the main text and in Eqn.~\eqref{hefflarge} and Fig.~\ref{figSehala} involve only $\sigma_x$ and $\sigma_z$ Pauli matrices. The pseudospin rotates in the XZ plane, and the winding is therefore along the Y axis. In the case of the complex Hamiltonian, shifting away from the collision point makes the $\sigma_y$ Pauli matrix appear in the decomposition of the effective Hamiltonian (unfortunately, the analytical expressions are too cumbersome and the effect was described numerically), and the winding vector therefore rotates away from the Y axis.

This rotation continues, until the winding vectors meet again in a certain plane, but pointing in the opposite directions, at the moment of the annihilation. The topological considerations from the main text guarantee that the rotation of the winding vectors in this picture has exactly the right angle ($\pi/2$ for each of the winding vectors).


\subsection{Additional experimental results}
\begin{figure*}
    \centering
    \includegraphics[width=1.0\linewidth]{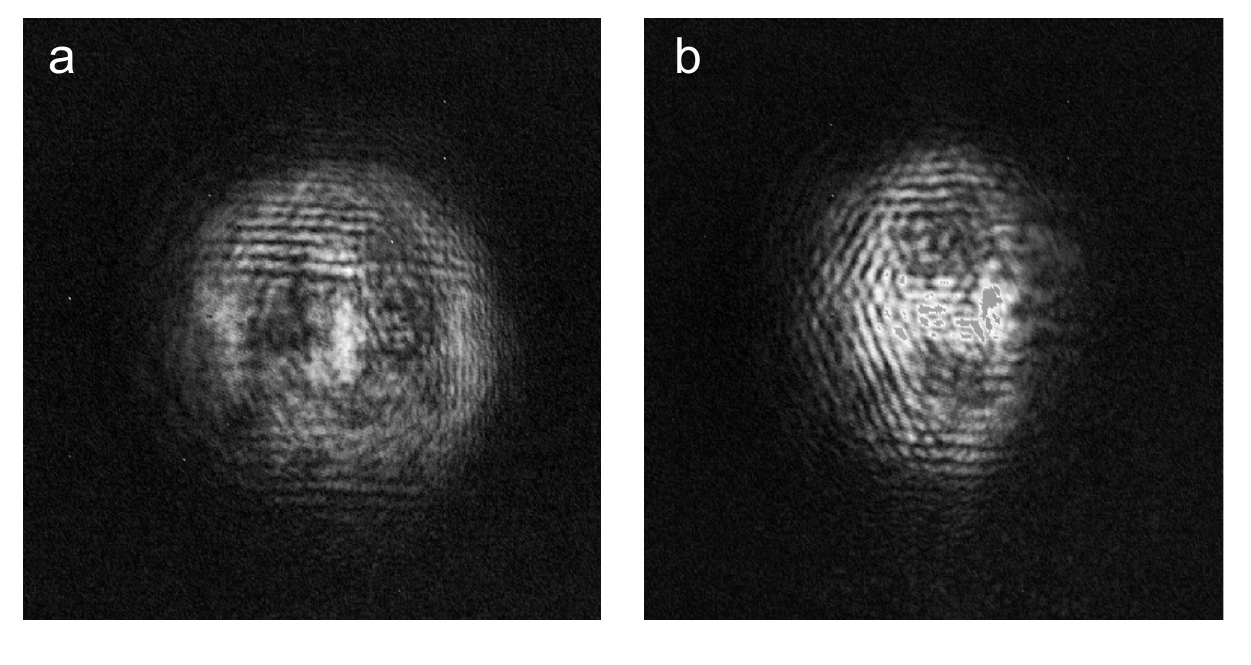}
    \caption{\textbf{Unfiltered images of conical diffraction for Fig. 2.} \textbf{a}~Original image for Fig.~2c, $E_A<0$. \textbf{b}~Original image for Fig.~2g, $E_A>0$.}
    \label{figS}
\end{figure*}

\begin{figure*}
    \centering
    \includegraphics[width=1.0\linewidth]{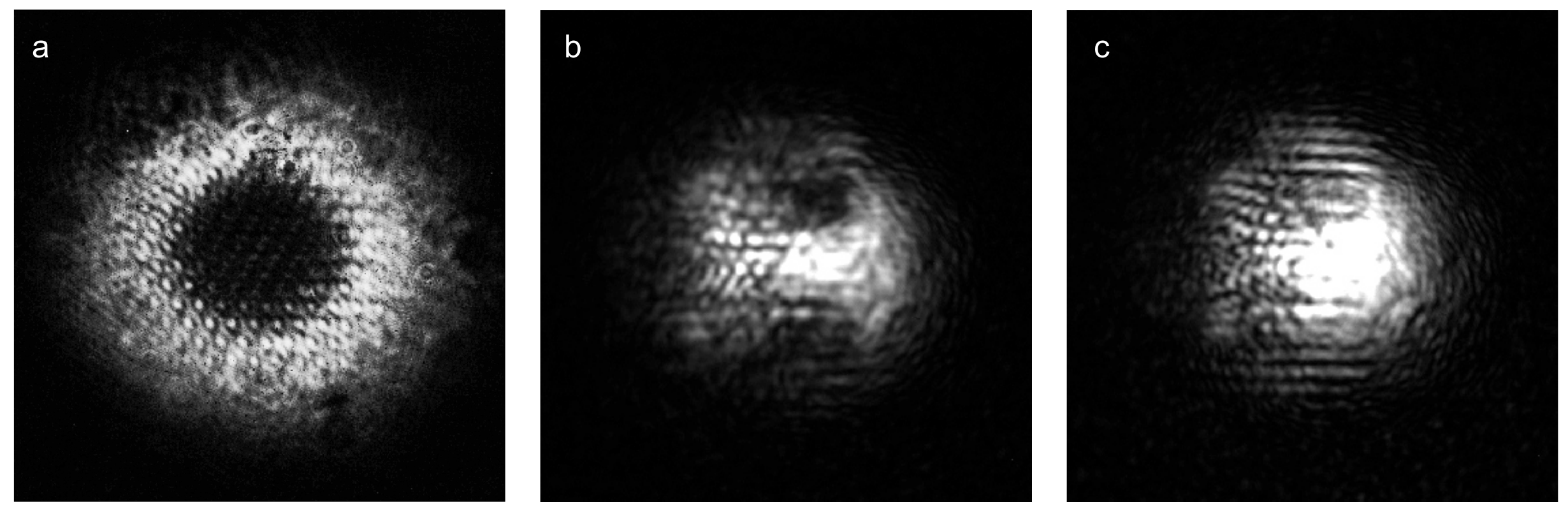}
    \caption{\textbf{Unfiltered images of conical diffraction for Figs. 1 and 4.} \textbf{a}~Original image for Fig.~1f (conical diffraction). \textbf{b}~Original image for Fig.~4a (before annihilation).
    \textbf{c}~Original image for Fig.~4c (after annihilation).
    }
    \label{figS4}
\end{figure*}

In this section we provide the images of the conical diffraction presented in the main text (Figure 2) without filtering. Figure~\ref{figS} shows the two images corresponding to panels 2c and 2g of the main text. The Dirac points are still visible as overall minima of intensity, but now they are more difficult to see because of the modulation due to the lattice.
Figure~\ref{figS4} presents three images corresponding to panels 1f, 4a, and 4c of the main text. A single Dirac point is well visible in ordinary conical diffraction. Two Dirac points are also quite visible, as well as their absence in the last panel.

\end{document}